# Wave Climate from Spectra and its Connections with Local and Remote Wind Climate

Haoyu Jiang [1,2,3], Lin Mu [1,3*]

[1] College of Marine Science and Technology, China University of Geosciences, Wuhan, China

[2] Laboratory for Regional Oceanography and Numerical Modeling, Qingdao National Laboratory for Marine Science and Technology, Qingdao, China

[3] Shenzhen Research Institute, China University of Geosciences, Shenzhen, China

Corresponding Author: Lin Mu (Moulin1977@hotmail.com)

# ABSTRACT

Because wind-generated waves can propagate over large distances, wave spectra from a fixed point can record information about air-sea interactions in distant areas. In this study, the spectral wave climate is computed for a specific location in the tropical Eastern Pacific Ocean. Several well-defined partitions independent of each other, referred to as wave-climate systems, are observed in the annual mean wave spectrum. Significant seasonal cycling, long-term trends, and correlations with climate indices are observed in the local wave spectra, showing the abundant climatic information they contain. The projections of the wind vector on the direction pointing to the target location are used to link the spectral wave climate and basin-scale wind climate. The origins of all the identified wave climate systems are clearly shown in the wind projection maps and some are thousands of kilometers away from the target point, demonstrating the validity of this connection. Comparisons are made between wave spectra and the corresponding local and remote wind fields with respect to seasonal and interannual variability, as well as the long-term trends. The results show that each frequency and direction of ocean wave spectra at a certain location can be linked to the wind field for a geographical area from a climatological point of view, implying that it is feasible to reconstruct a spectral wave climate from global observational wind field data and wind climate monitoring using observations of wave spectrum geographically far away.

# 1. Introduction

Wind-generated surface gravity waves (simply called waves hereafter) are a fundamental and ubiquitous phenomenon at the air-sea interface. They impact many aspects of human life, from industrial activities such as seafaring and port operations, to recreational activities like surfing and yachting, and play a crucial role in many geophysical processes such as momentum exchange at the air-sea boundary layer. The studies of wave climate are important both from societal and scientific perspectives, thus, are widely conducted in different scales, including global (e.g., *Young* 1999; *Chen et al.* 2002; *Hemer et al.* 2010, 2013; *Semedo et al.* 2011, 2013; *Young et al.* 2011, 2012; *Fan et al.* 2012, 2013, 2014), basin (e.g., *Gulev and Hasse* 1998; *Gulev and Grigorieva* 2006; *Stopa and Cheung* 2014; *Liu et al.* 2016), regional (e.g., *Weisse and Günther* 2007; *Anoop et al.* 2015; *Semedo et al.* 2015), and local point scales (e.g., *Bromirski et al.* 2005; *Gemmrich et al.* 2011; *Espejo et al.* 2014; *Portilla et al.* 2015a, 2015b, 2016).

Waves are generated by wind but are not always coupled to local winds because they can propagate away from their origins and become swells. Swells can propagate over thousands of kilometers with little energy loss (e.g., *Snodgrass et al.* 1966; *Ardhuin et al.* 2009; *Jiang et al.* 2016). Therefore, for a given point in the ocean, the sea state might be the superposition of a local wind-sea system and more than one swell system originated from remote regions. Analyzing waves using mean spectral parameters such as significant wave height (SWH), mean wave direction (MWD), and mean wave period (MWP), only provides a limited description of the wave field and might be misleading in a mixed sea state. Therefore, significant efforts have been made on the characterization of waves using spectra, such as spectral partitioning schemes (e.g., *Gerling* 1992; *Wang and Hwang* 2001;

*Hanson and Phillips* 2001; *Portilla et al.* 2009; *Hwang et al.* 2012). Wave spectral information and partitioning schemes are receiving increasing attention in both the scientific community and marine industries, as they can reveal the contributions of each wave system (e.g., *Portilla et al.* 2013; *Portilla and Cavaleri* 2016; *Portilla* 2018).

Waves carry the information of the wind generating them. Thus, the local wave climate derived from wave spectra can shed some information on both local and far-field wind climates (*Portilla et al.* 2016). However, only a few studies about the spectral wave climate have been conducted mainly because of two reasons: 1) Buoys with the ability to measure wave spectra have not accumulated data for 30 years which is the minimum for climate studies as recommended by the World Meteorological Organization. 2) Although methods for reconstructing wave spectra from the four Fourier coefficients measured by buoys, such as the maximum entropy method (*Earle et al.* 1999), are widely applied in engineering, they are also known to have problems such as reducing the directional spread and generating spurious peaks. *Bromirski* et al. (2005) related the energy in different frequency bands of several North Pacific buoys with the sea level pressure (SLP) modes using principal component analysis and found some climate signals in waves. *Pérez et al.* (2014) presented a method for evaluating the source and travel time of the wave energy reaching any location in the open ocean using global wave (reconstructed) spectral information. *Espejo et al.* (2014) reconstructed a long-term spectral wave climate from the SLP field and relatively short-term buoy data using cluster analysis and characterized the spectral wave climate in the northeast Atlantic. *Portilla et al.* (2015a, 2015b) built a systematic methodology for analyzing the local wave climate based on the spectral-domain probability density function of partitioned peak wave periods and peak wave directions, and an atlas of global spectral wave climate (GLOSWAC) is developed based on this method (*Portilla* 2018). They showed that El Niño events have a plausible signature on the wave spectra in the eastern equatorial Pacific (*Portilla et al.* 2016), implying that there could be abundant climatic information in wave spectra.

The aims of this study are to establish a methodology for describing the local spectral wave climate and provide some insights into the potential connection between the local spectral wave climate and the local/remote wind climate. A swell-dominated location in the Tropical Eastern Pacific Ocean is selected as an example to illustrate how local spectral information is correlated with the wind information from thousands of kilometers away. The rest of the paper is as follows: In section 2, the data of wave spectra and sea surface wind used in this study are introduced, and some climate indices are described. Section 2 also presents the methodology for analyzing spectral wave climate and connecting it with wind information. In section 3, the climate and the variability of wave spectra at a selected location are detailed with a discussion on how they are connected to the local/remote wind climate. Section 4 is a summary of major findings and concluding remarks.

## 2. Data and methodology

*2.1 Data of wave spectra and wind field*

Contemporary numerical wave models can provide a reliable spectral description of waves, especially in the open ocean, and wave spectra from them are consistent with those measured by buoys (e.g., *Stopa et al.* 2016; *Portilla et al.* 2015b). Due to the two problems with observational wave spectra mentioned earlier, numerical wave models are regarded as useful tools to investigate the wave climate (e.g., *Semedo et al.* 2011; *Stopa and Cheung* 2014; *Portilla et al.* 2016). This study uses a 39-year (1979-2017) record of directional wave spectra and 10-m wind field data from the European Centre for Medium-Range Weather Forecasts (ECMWF) Reanalysis-Interim (ERAI) (*Dee et al.* 2011). ERAI is a coupled atmosphere and surface waves reanalysis covering the period from 1979 to the present. The horizontal resolution of the atmospheric model is approximately 79 km on a reduced Gaussian grid and the resolution of the coupled wave is approximately 110 km. The wave spectral information from ERAI can be downloaded with a temporal-spatial resolution of 6 h × 1 ° × 1 ° using ECMWF Web API. Each spectrum at a certain time and location is divided into 30 frequency bins that increase exponentially from 0.0345 to 0.5473 Hz and 24 directional bins with 15 ° spacing.

The location selected in our case study is a point in the Pacific Ocean at 5 °N, 120 °W (hereafter, Point X). This is a swell-dominated position with a relatively complex wave condition located in both the Pacific "swell pool" (*Chen et al.* 2002) and Pacific "crossing swell pool" (*Jiang et al.* 2017a) where several wave partitions often coexist at the same time. The swells generated by westerlies in both hemispheres can propagate into this point while it is also impacted by the Inter-Tropical Convergence Zone (ITCZ)-induced trade winds. Ignoring refraction, reflection, and diffraction during wave propagation, and assuming that waves propagate along great circles on the ocean surface, the regions where wave energy with appropriate directions can impact the selected location can be estimated by drawing great circles in different directions from the selected location, as shown in *Pérez et al.* (2014). The propagating distances and the azimuth from and to the selected location are shown in Figures 1a, 1b, and 1c, respectively. In Figures 1b and 1c, 0 ° and 90 ° refer to the azimuth pointing to the north and east, respectively. The figures can be interpreted as the following: when the wave energy at a given location propagates in the direction of Figure 1(c), it will impact the wave energy in the opposite direction of Figure 1(b) at the selected location after propagating over the distances in Figure 1(a). Therefore, it can be deduced from Figure 1 that waves in the tropical Indian Ocean propagating in the direction of ~140° might theoretically propagate into the target point. However, the frequency dispersion, angular spreading, and the limited but still significant swell dissipation will attenuate the wave energy along propagation (*Jiang et al.* 2017b), resulting in the swell energy coming from small and distant regions hard to be detected. As waves are generated by wind, Figure 1 also implies that the wave climate at a single point in the open ocean has the potential to partially reflect the wind climate at a larger scale via information contained in swells.

*2.2 Climate indices*

According to *Stopa and Cheung* (2014), the wind and wave climates in the Pacific Ocean are impacted by climate oscillations, including the El Nino-Southern Oscillation (ENSO), the Antarctic Oscillation (AAO), and the Arctic Oscillation (AO). Many indices are developed to describe these climate oscillations. The ENSO has a strong signature in the equatorial Pacific which is usually indicated by atmospheric pressure or temperature anomalies across the Pacific basin. A frequently-used index for the ENSO is the Southern Oscillation Index (SOI) which is defined as the difference of the SLP anomalies between Tahiti and Darwin. The AAO, also known as the Southern Annular Mode, is defined as a low pressure surrounding Antarctica that moves north or south as its mode of variability (*Gillett et al.* 2006). The AO (Northern Annular Mode) is an oscillation equivalent to the AAO in the Northern Hemisphere (NH) (*Thompson and Wallace* 1998). The indicators of these two oscillation are the AAO Index (AAOI) and the AO Index (AOI) which are both defined by the first empirical orthogonal function of geopotential height. In this study, these three climate indices are employed to identify the correlations between the local spectral wave climate and climate oscillations. Monthly SOI, AAOI, and AOI data are downloaded from National Oceanic and Atmospheric Administration Earth System Research Laboratory (https://tinyurl.com/yd2nhcu2).

*2.3 Spectral wave climate*

The global wave energy can be described as 5-dimensional (5D) wave spectral densities $G(t, \varphi, \lambda, f, \theta)$ with the given time $t$, latitude $\varphi$, longitude $\lambda$, frequency $f$, and direction $\theta$. Most studies focus on the parameters integrated along $f$ and $\theta$ such as SWH ($H_{m0}$), MWD ($T_{m-1,0}$), and MWP ($\theta_m$) where:

$$H_{m0} = 4.04\sqrt{\iint f^0 G(f,\theta) df\, d\theta} \qquad (1)$$

$$T_{m-1,0} = \frac{\iint f^{-1} G(f,\theta) df\, d\theta}{\iint f^0 G(f,\theta) df\, d\theta} \qquad (2)$$

$$\theta_m = \arctan\left[\frac{\iint \sin(\theta) G(f,\theta) df\, d\theta}{\iint \cos(\theta) G(f,\theta) df\, d\theta}\right] \qquad (3)$$

In this case, the 5D array is simplified into a 3D temporal-spatial array $H(t, \varphi, \lambda)$ where $H$ can be any integrated parameters ($H_{m0}$, $T_{m-1,0}$, or $\theta_m$). Data processing methods are developed based on this 3D array, from the simple climatological mean to classical statistical techniques such as empirical orthogonal functions. When $\varphi$ and $\lambda$ are fixed, $G(t, \varphi, \lambda, f, \theta)$ also becomes a 3D array $G(t, f, \theta)$. Each $<f, \theta>$ corresponds to a time series of spectral densities, like each $<\varphi, \lambda>$ corresponding to a time series of SWH in $H(t, \varphi, \lambda)$. Some of these time series of spectral densities at Point X are shown in Figure 2 using 20-year ERAI wave spectra over 1981-2000 (with the data averaged monthly for clarity). Annual cycles are observed in most of the series with the energy propagating in different directions in different phases. For instance, the wave energy in the directions of 127.5 ° and 232.5 ° (a direction of 0 °/90 ° corresponds to the wave energy propagating towards the North/East) is in the same phase which generally reaches the maximum in boreal winter, while the wave energy in the other two directions has the opposite phase which generally reaches the maximum in boreal summer. Some exceptionally high peaks, such as the peaks at <0.05Hz, 127.5 °> and <0.13Hz, 232.5 °> in the boreal winter of 1982/1983 and 1997/1998, are well-defined. They can be regarded as the signatures of ENSO on the ocean wave spectra and indicate that winter storms and trade winds in the NH are stronger during El Niño events. These features show that the information derived from local wave spectra is rich from a "climatological" point of view.

The 6-hourly wave spectra at the selected point are averaged to obtain the annual mean wave spectra (AMWS) and the seasonal mean wave spectra (SMWS). The four seasons are organized as December–February (DJF), March–May (MAM), June–August (JJA), and September–November (SON). The standard deviations (STD) and long-term trends for each spectral bin are computed to a spectrum of STD (SSTD) and a spectrum of trend. In order to analyze the relationship between the local spectral wave climate and climate oscillations, the correlations between the time series of energy for each spectral bin and the aforementioned climate indices are computed, resulting in "directional spectra" of correlation coefficients.

*2.4 Wind projections and relationship to wave spectra*

The evolution of wind-wave energy is a complex process and is represented by the wave action balance equation in numerical wave models. From a statistical point of view, however, the distribution of spectral wave energy at a given location is determined by the wind field in a certain region. For instance, the regions with potential impacts on the wave spectra at Point X are shown in Figure 1. However, it is noted that not all the wind in these regions can leave their signals on the wave spectra at that location. For instance, winds blowing in the directions opposite to the directions in Figure 1c will not generate any wave energy that can reach Point X. Considering that wind-generated wave energy is positively correlated with the wind component in the wave propagation direction, the projection of the wind vector on the direction pointing to the target point (hereafter, wind projection) is employed to link the basin-scale wind field with the local wave spectra:

$$U_{proj} = MAX\left[0, U\cos(\theta_U - \theta_A) \cdot \delta\right] \qquad (4)$$

where $U$ is the 10-m wind speed, $\theta_U$ is the direction of the wind, $\theta_A$ is the azimuth direction from a geographical location to the target point, and $\delta$ is the land blocking factor which equals 0 (1) if the great circle between the location and the target point is (not) blocked by land. For any given point in the ocean such as Point X in this study, the corresponding wind projections at different geographical locations also make up a 3D array $U_{proj}$ ($t$, $\varphi$, $\lambda$) covering the global ocean.

The normalized cross-correlation array between the local wave spectra and the corresponding wind projection array is calculated for Point X to illustrate the effectiveness of the wind projection:

$$r(\tau, f, \theta, \varphi, \lambda) = \frac{\sum_{t=1}^{n}\left[G(t-\tau, f, \theta) - \overline{G}(f, \theta)\right]\left[U_{proj}(t, \varphi, \lambda) - \overline{U_{proj}}(\varphi, \lambda)\right]}{\sqrt{\sum_{t=1}^{n}\left[G(t-\tau, f, \theta) - \overline{G}(f, \theta)\right]^2}\sqrt{\sum_{t=1}^{n}\left[U_{proj}(t, \varphi, \lambda) - \overline{U_{proj}}(\varphi, \lambda)\right]^2}} \qquad (5)$$

where $\tau$ is the lag time between the spectral series and wind projection series which is introduced here because the wave generated by the remote wind takes time to propagate to the target point. At latitude $\varphi_n$ and longitude $\lambda_n$, the values of $r$ ($\tau$, $f$, $\theta$; $\varphi_n$, $\lambda_n$) can be interpreted as a 3D matrix of correlation coefficients between the wind projections and the spectral densities in all spectral bins ($f$ and $\theta$) with different time differences ($\tau$). The maximum value of $r$ ($r_{max}$) is therefore regarded as the correlation value between wind projection at $<\varphi_n, \lambda_n>$ and the wave spectra at Point X. The $\tau$ corresponding to $r_{max}$ can be used to define the average travel time for generated waves to propagate from different locations to Point X while another definition of average travel time of wave is given by *Pérez et al.* (2014). Figure 3a shows the spatial distribution of correlations between the wind projections and the wave spectra at Point X, and the values of $\tau$, $f$, and $\theta$ that correspond to $r_{max}$ are shown in Figures 3b, 3c, and 3d, respectively.

The garden sprinkler effect-like pattern in Figure 3a and the discontinuous edges in 3b, 3c, and 3d are due to the discreteness of spectral frequency and direction bins, which can be reduced using the spectra with a higher resolution. Unsurprisingly, the highest $r_{max}$ is observed in the region close to Point X with the highest values of more than 0.9. In this region, the time lag between the wind and wave is generally within 50 h and the wind projection is correlated with wind-sea energy at frequencies higher than 0.15 Hz. However, it is noted that there is no strict definition of when wind-seas turn into swells. Except for the wind exactly at Point X, the wave energy generated anywhere in the ocean, no matter how close or far, need some time arrive at Point X, leading to the time lag always being larger than zero. Outside this region, the maximum correlations between the wind projections and the wave spectra at Point X are relatively lower, but statistically significant correlations (with a $P$ value of 0.01) are found over almost all of the Pacific Ocean except in the tropical central Pacific. The correlation coefficients are greater than 0.6 in the trade-wind zone to the southeast of Point X and are greater than 0.5 in the storm track regions at midlatitudes and the tropical west Pacific, which are thousands of kilometers away. The $r_{max}$ in these regions preliminarily corresponds to relatively low frequencies, and the corresponding time lags (travel time for the wind signals) generally increase with distances. Whether near or far, the corresponding spectral directions (Figure 3d) show a pattern of rotation around Point X, which is in line with the azimuth directions in Figure 1b. These results indicate that the signals of the wind field, both local and remote, can be recorded by local wave spectra, and the spectral densities of each frequency and direction can be statistically linked with the wind projections from a certain geographical region. Therefore, the wind projection can be employed as the parameter to describe the connections between wind climate and spectral wave climate. The annual mean, seasonal mean, long-term trends, and the correlations with climate indices of $U_{proj}$ ($t, \varphi, \lambda$) are computed to evaluate the climatological relationship to wave spectra at Point X.

## 3. Results and Discussion

*3.1 Linking spectral wave climate with wind climate*

The AMWS of the selected location are displayed in Figure 4a. There are four energy peaks at <0.07Hz, 20 °>, <0.07Hz, 135 °>, <0.14Hz, 240 °>, and <0.12Hz, 320 °>, which correspond to four wave climate systems. Here, these four systems are labelled as System A, B, C, and D, respectively. To analyze whether these systems are wind-sea- or swell-related, the correlation coefficients between local wind projections and the spectral densities in each spectral bin are computed, and the results are shown in Figure 4b. The azimuth direction $\theta_A$ in Equation 4 is not applicable for local winds, thus it is defined as the direction in the spectrum. This correlation coefficient can be employed as an indicator to describe whether the sea state is wind-sea-dominated or swell-dominated from a climatological point of view: a high correlation between wind speed and wave energy means the waves are strongly coupled to the local wind so that it corresponds to wind-seas, and vice versa for swells. As expected, although apparent anisotropy can be observed in the distributions of correlation coefficients, the high-frequency part of the spectrum generally has relatively high correlations with the local winds. Systems A and B have low frequencies and sharp energy peaks, and their corresponding correlations with the local winds are less than 0.1, which means that most of their energy is swell-related coming from far fields. The correlations between local winds and most spectral bins of Systems C and D are greater than 0.5, and these two systems have higher frequencies and wider directional spreads so that they are more wind-sea-related than Systems A and B. However, it is also noted that there is no strict distinction between wind-seas and swells. The low-frequency components of Systems C and D also show low correlations of less than 0.3, indicating that some of the energy in these two system is also swell-related. The AMWS shows that there could be more than one wind-sea-related system and two swell-related systems coexisting in the wave climate at the same point. The values of $H_{m0}$, $T_{m-1,0}$, and $\theta_m$ corresponding to the AMWS are 2.13 m, 10.4 s, and 14.5 °, respectively. These integrated parameters cannot adequately describe the wave state from a climatological point of view even if wind-seas and swells are treated separately.

The SSTD (Figure 4c) shows the degree of scattering of spectral densities in each frequency and direction bin. The pattern of the SSTD is similar to that of the AMWS, which means the frequency and direction with higher mean wave energy usually also have higher variability of wave energy. Meanwhile, there are also some differences between the two patterns. The AMWS of System A is more energetic than that of System B in Figure 4a, but the SSTD of System B is larger than that of System A in Figure 4c. A harmonic analysis is applied to the time series of the spectra using the Fast Fourier Transform to extract the amplitude of annual cycle for each frequency and direction bin, with results shown in Figure 4d. The annual amplitude of spectral density is much larger for System B than System A, showing that the larger SSTD of System B is primarily due to the more intense seasonality of the swells coming from the NH than those from the Southern Hemisphere (SH). Both of the wind-sea-related peaks, C and D, are observed in lower frequencies than those in the AMWS pattern. The AMWS and SSTD are computed at some other points (not shown here), and this phenomenon is well-defined for nearly all wind-sea systems, indicating that spectral densities of wind-sea-related systems are usually more scattered in the low-frequency parts, but the reason for this is unknown at this stage. Another noteworthy feature is that the SSTD values are comparable to and sometimes greater than the AMWS values in most of the spectral bins (the maximum SSTD is larger than the maximum AMWS). Because the values of spectral density are never less than zero, the relatively large STDs mean that the distributions of spectral densities generally have high skewness and long tails, which is confirmed by the probability density function (not shown here).

Since wind projections and wave spectra are correlated, the climatology of wind projections should also be in line with the climatology of wave spectra. The spatial distributions of the annual mean and the STD of the wind projections for Point X are displayed in Figures 5a and 5b. For all the wave climate systems in Figure 4, their respective origins are clearly shown in Figure 5a. The origins of swells in different part of the global ocean have been identified by *Alves* (2006), which are in line with the results here. Meanwhile, the method described in this paper is able to identify the wave energy source for a given target point, which provides a different perspective. The main sources of wave energy include the westerlies and trade wind zones in both Hemispheres. Regions with high wind projections in the westerlies in the SH and NH are the sources of the energy in Systems A and B in Figure 4a, respectively. Although the distance from Point X to the source region of System B is a bit larger than the distance to the source region of System A, the wind projections are significantly larger in the SH, resulting in the mean energy of System A being higher than System B. The trade winds along South America which correspond to System D in Figure 4a is another important source region for waves at Point X, which are more well-defined than the trade winds in the NH in Figure 5a. For this reason, System D is more well-defined than System C in Figure 4a. Another noteworthy feature in Figure 5a is the wind projections for the California low-level coastal jet (CLCJ, *Burk and Thompson* 1996) which are clearer than the projections of the trade winds in the NH. However, the corresponding wave system of CLCJ is not directly observed in the AMWS because the direction of the wave energy from CLCJ to Point X overlap with the wave energy from the westerlies in the NH. In GLOSWAC (*Portilla* 2018, the product is available at: http://www.modemat.epn.edu.ec/nereo/), the wave system corresponding to CLCJ (hereafter System E) is well-defined in the long-term probability density distribution of wave partitions (PDDWP). As pointed out by *Portilla et al*. [2015b], the AMWS is usually smoother than the PDDWP so that not all single wave systems, especially those with relatively low energy, are recognizable in the AMWS. Meanwhile, the signature of System E is still vaguely observed from the asymmetry of the spectral densities near

System B: Due to the energy from System E, the right side of System B along the propagation has more energy than the left side.

The STD of wind projections show a pattern similar to the mean wind projections. The largest variability of wind projections is found in the westerlies of both Hemispheres with similar values of STD, corresponding to the high SSTD of Systems A and B. In this case, the source region of System B is closer to Point X than that of System A, which explains System B's higher values because a larger distance means large energy attenuation due to the angular spreading effect. The variability of CLCJ is also significant, but its signature in the SSTD is also partially overwhelmed by System B as in AMWS. The trade winds in both Hemispheres in Figure 5b are still observable but not as clear as they are in Figure 5a, which is also in line with the relatively vague signals of Systems C and D in the SSTD of Point X.

*3.2 Seasonal variability*

The SMWS at the selected point is displayed in Figure 6. The two swell Systems A and B in Figure 4a show opposite phases peaking in JJA and DJF, respectively. System A which originates from the extratropical storm track in the SH is well-defined all over the year, but its energy is significantly lower in DJF than in the other three seasons, which is because wind storms in the SH are the weakest in DJF. System B shows a stronger seasonality (as shown in Figure 4d) with the highest wave energy in DJF while it almost disappears in JJA when the swells from the westerlies in the North Pacific cannot propagate to Point X. Meanwhile, System E which is generated by the CLCJ is clearly identified at <0.12Hz, 170°> in JJA as shown in Figure 6c. The two more wind-sea-related systems, C and D, also show opposite phases with respect to seasonal variability. These two systems are generated by the trade winds from different hemispheres, which are stronger in the corresponding hemispheric winter due to the seasonal shift of the ITCZ. System C is clearly observed in DJF and is still identifiable from the outline of the contours in MAM, but it almost disappears in JJA and SON, indicating that the northeast trade winds in the NH generally cannot impact the wave state at Point X during the boreal summer and autumn. System D is observed in all four seasons due to the potential long fetch of southeast trade winds to Point X even in DJF and MAM, while its energy is significantly higher in JJA and SON. Although clear seasonality of the spectra is observed for all wave climate systems, the seasonality of SWH is not as clear due to the cancellation effect of the phase-opposite variations of these systems: The mean total SWH at Point X is the highest in DJF and the lowest in JJA with a difference of only 0.23m.

Seasonal mean wind projections for the Point X are displayed in Figure 7. In line with the SMWS of System A in Figure 6, the seasonality of the wind projections in the westerlies of the SH is relatively low. On one hand, the maximum values of wind projection in this region are similar across all four seasons. On the other hand, the extent of the fetch in DJF is the smallest among the four seasons, leading to the least energy propagating into Point X during that time. It is noted that the wind speed in the westerlies in the Southern Ocean to the south of Australia and New Zealand should be significantly higher in JJA than in DJF, but the variability of wind speed are mainly in the component perpendicular to the direction to Point X [Figure 1 of *Semedo et al.* (2011)]. Thus, the seasonal variability of wind speed in this region has little influence on the energy of System A. The maximum values for trade wind projections along South America are ~6 m/s in DJF and MAM, and ~7 m/s in JJA and SON. Meanwhile, the ITCZ moves to a position to the north of the equator in JJA and SON so that the fetch of trade winds is also closer to Point X in JJA and SON than in DJF and MAM (Figure 7). Both reasons explain the seasonal variability of System D in Figure 6. In contrast, the seasonality of the wind projections in the NH is much greater. The seasonal shift of the ITCZ makes the trade wind projections in the NH more clear in DJF and MAM but become vague in JJA and SON, which is also in good agreement with the seasonal variability of System C. Reaching its maximum in DJF, the wind projections in the westerlies of the NH exhibit strong seasonality in agreement with the seasonal variability of System B. In JJA, the signature of the westerlies in the NH is not observable while the wind projections of the CLCJ are the strongest, resulting in the disappearance of System B and the appearance of System E in this season. Another feature of the NH westerlies with respect to the wind projections is the northward shifting during SON in Figure 7, which corresponds to the shift of the spectral peak of System B from ~135 ° in DJF to ~150 ° in SON (Figure 6). For regions where the wind projections are small in magnitude but highly correlated to the spectra in Figure 3, the impacts of the wind climate on the spectral wave climate are not directly observable with respect to seasonal variability. However, such impacts might be distinguished from the interannual variability as shown in the following.

*3.3 Interannual variability*

To analyze the link between the local spectral wave climate and large-scale climate oscillations, correlation coefficients are computed between the anomalies of monthly-averaged spectral densities and the three universally accepted climate indices described in section 2 for all spectral grids (Figure 8). The anomaly of monthly average is defined as:

$$A_{y,m}(f,\theta) = \overline{G_{y,m}}(f,\theta) - \overline{G_m}(f,\theta) \quad (6)$$

to eliminate the impact of seasonal variability where $\overline{G_{y,m}}$ is the monthly mean of year $y$ and month $m$ and $\overline{G_m}$ is the mean of month $m$ for all years. The corresponding correlation coefficients of monthly-averaged wind projection anomalies with the climate indices are displayed in Figure 9.

Significant correlations are observed between the wave spectra and the climate indices as well as between the wind projections and the climate indices. In Figure 8a, the highest correlation with SOI ($r \approx -0.5$) is observed near <80°, 0.13 Hz>, which is a swell-related spectral regime because the correlation for this region in Figure 4b is less than 0.3. Significant correlation coefficients of ~±0.4 are found in the wind-sea-related high-frequency regime of the spectrum with opposite phasing between the northward and southward directions. The regions where the five identified wave climate systems are located are not well-defined in Figure 8a, and the energy of Systems A, D, and E is not significantly correlated with the SOI. Meanwhile, the regime corresponding to the peak of System B significantly correlates negatively with the SOI ($r \approx -0.4$). In Figure 9a, a well-marked feature is the strong negative correlation ($r \approx -0.7$) in the western equatorial Pacific, which appears to correspond to the strong negative values in Figure 8a. Although the predominant winds are westward in this region, west winds also occurs sometimes especially in DJF as shown in Figure 7a, and the waves generated by them are able to reach Point X as shown in Figure 12 of *Alves* (2006). The correlation coefficients between the wind projections in the western equatorial Pacific and the wave energy in the aforementioned spectral regime can reach up to 0.7 (Figure 3), also demonstrating that the eastward wave energy is able to propagate to Point X. Therefore, the strengthened/weakened west winds in this region during the El Niño/La Niña events leads to stronger/weaker eastward wave energy at Point X, resulting in the negative correlations with the SOI. The other spectral regions with significant correlation coefficients in Figure 8a can also find their sources in Figure 9a: The significant correlation in the high frequency regime of Figure 8a is in agreement with the dipole-like pattern of the wind projections near Point X, showing that the Southward/Northward component of the local winds is stronger during the El Niño/La Niña event. Besides, the El Niño/La Niña event corresponds to stronger/weaker wind in the North Pacific westerlies, which is in line with the negative correlations between the energy anomalies of System B and SOI.

However, not all the regions with significant correlation coefficients with SOI in Figure 9a can find their signatures in Figure 8a. For instance, the trade winds along the coast of South America are weaker during El Niño years due to the anomalies of west wind, which also lead to weaker wind projections to Point X. Thus, most of the trade wind zone in this region is positively correlated with the SOI. Meanwhile, the mean position of the ITCZ moves southward during El Niño years, causing the north part of this trade wind zone to have negative correlations with the SOI. Due to the cancellation of the northern part of the trade winds, the negative correlations between the spectrum and the SOI are not significant in the energy peak of System D. There are also the two regions with significant but weak positive correlation coefficients ($r \approx$ -0.2) in the high latitudes of both hemispheres in Figure 9a. After the propagation of large distances, however, the wave energy generated by them only has insignificant positive correlations with the SOI in low-frequency regime around 10° and 170°.

The correlation values of the wave spectra at Point X with the AAOI and AOI are generally lower than with the SOI, and significant correlations are mostly found at low frequencies dominated by swells. For AAOI, significant correlations are only observed in the northern half of the spectrum with the largest positive correlations ($r \approx 0.35$) near the peak of System A, while significant negative correlations are found on both sides of this regime. This low-high-low pattern is in good agreement with the wind projections to the south of 30°S in Figure 9b, although the correlation coefficients are significantly lower in Figure 8b after the long propagation of wave energy. When the AAO is in the positive phase, the westerly wind belt intensifies and contracts towards Antarctica. Stronger westerlies correspond to higher/lower wind projections to Point X to the west/east of 120°W. Thus, the wind projections in the westerly wind belt also shows strong positive/negative correlations ($r \approx \pm 0.65$) with the AAOI to the west/east of 120°W. Meanwhile, the wind projection in the north of the westerly wind belt shows a significant negative correlation with AAOI due to the shifting of the westerly wind belt with the phase of AAO.

Symmetry can be observed between the correlations of the wave spectra with AOI and those with AAOI. Significant correlations locate only on the southern half of the spectrum for AAOI in Figure 8c, especially in the southeast quadrant, indicating that the AAO and AO mainly impact the swells propagating northward and southward, respectively, in the tropical Pacific Ocean. Significant negative correlations ($r \approx -0.3$) are found near the peak of System B, and positive correlations are found in the low-frequency part of 180°. This pattern is also in good agreement with Figure 9c where AO's impact on the wind projections is mostly reflected in the regions to the north of 30°N. The AO's impact on the westerly wind belt in the NH is similar to the AAO's in the SH: The westerly wind belt intensifies and contracts towards the polar region in the positive phase of AAO. However, due to the presence of the North American continent and the Aleutian Islands, the area in Figure 9c with positive correlation coefficients with AOI is much smaller than the corresponding region in Figure 9b. Besides, the correlation values are lower ($r \approx 0.3$) in Figure 9c than in 9b, resulting in its relatively small signatures on the spectrum of correlation (Figure 8c). On the contrary, the area with negative correlation with AOI corresponding to the shifting of the westerly wind belt with the phase of AO is larger, which leads to a larger area of negative correlations in Figure 8c.

*3.4 Long-term trends*

Due to the increase in global sea surface wind speeds, the global mean SWH is observed to increase over the last several decades (e.g., *Aarnes et al.* 2015; *Young et al.* 2011; *Hemer et al.* 2010). Wind and SWH trends are different across different regions and are not uniformly positive because of the inhomogeneity of wind speed increase and the propagation of swells. Meanwhile, the wave climate at a given location is composed of different wave systems which might have different trends of wave energy. The distributions of linear spectral density trends at Point X, and corresponding wave projection trends, are shown in Figures 10a and 10b, respectively. The Mann-Kendall method is employed to calculate the trends and test their significance. It is noted that the use of reanalysis to describe trends is problematic because of the change in the quality and quantity of remote sensing observations assimilated into the data. Particularly, more observations are available from altimeters from 1991 onwards, which produces spurious trends in the reanalysis data (*Aarnes et al.* 2015). Thus, only the trends over the period from 1992 to 2017 are computed here. During this period, the linear trend of SWH at Point X is not significant even at the 90% confidence level in the ERA-I data. However, all the four wave climate systems identified in Figure 4 have significant increasing or decreasing trends, which implies that the trends of different wave systems neutralize the overall trend into an insignificant one. The spectral regimes of the two swell-related Systems A and B show significantly negative trends while C and D show positive trends, and System E shows no significant linear trend. The corresponding SWH trends for Systems A, B, C, D are -1.7 cm/yr, -2.9 cm/yr, 1.0 cm/yr, 3.2 cm/yr, respectively.

The pattern of the wind projection trends are relatively complicated, but their connections with the spectral trends can still be partially identified. It is noted that the increase/decrease in the magnitude of wind projection does not necessarily mean the increase/decrease in wind speed, as the change of wind direction also impacts the wind projection. The most obvious upward trends of the wind projections are found in the trade wind zone along the coast of South America, which is in good agreement with the largest increase of wave energy in System D. The trade winds in the NH also generally have a moderate upward trend in line with the weak but significant increase of wave energy in System C. For the two swell systems with decreasing trends in wave energy, their corresponding source regions in the westerlies of both hemispheres have significant decreasing trends in wind projection. Besides, the CLCJ region and the equatorial West Pacific also have significant increasing and decreasing trends, respectively, in wind projection. However, their signatures on the wave spectra are not observable. There are two potential reasons for this: One is the cancellation effect along the route of wave propagation. For instance, the upward trends of wind projections in CLCJ might be canceled by the downward trends in the westerlies as they have similar arrival directions at Point X, while the downward trend in the equatorial West Pacific might be canceled by the upward trend of the eastward component of local winds. The other reason is that the data assimilation of winds and waves in the reanalysis is done independently although they are coupled in the model, which might also lead to some inconsistencies between the trends of wind and waves considering that the signal of these trends are relatively weak. It is noted that a time span of only 26 years is extremely short for linear trend calculation because the results are heavily impacted by inter-annual signals such as the ENSO cycle. Therefore, the results here are mainly regarded as evidence of the link between wind projections and wave spectra with respect to linear trends.

## 4. Summary and Concluding Remarks

As waves can propagate over large distances, the wave spectrum at any location might contain some information about air-sea interactions in geographically distant regions. In this study, the spectral wave climate and its variability at a selected point in the tropical Pacific Ocean are characterized using 39-year wave spectra over 1979-2017 from ERAI. Because the spectral densities of each frequency and direction can be roughly connected to the wind fields of a certain geographical region from a climatological point of view, a simple parameter is defined as the projection of wind on the direction pointing to a given location. The climatological wave spectrum at this point is constituted of five wave climate systems from different origins, which cannot be adequately described by integrated wave parameters even if wind-seas and swells are treated separately. The seasonal, interannual, and long-term variability of these wave climate systems are also independent of each other, implying that the local wave spectra have rich climatic information. The wind field data from ERAI illustrates that the generating areas of all wave climate system in the climatological spectrum are well-defined in the corresponding maps of wind projection climatology, confirming the effectiveness of this parameter. The wind-sea-related/swell-related energy in the higher/lower frequencies of the spectra can be connected with the projections of local/distant winds, and the footprints of corresponding seasonal and interannual variability and long-term trends of the wind climate are also found in the spectral wave climate. Therefore, the basin-scale wind climate and its variability are partially recorded by the wave spectra at a fixed point, and the wave spectra can be a potential indicator of the climate system.

The connection between the spectral wave climate at a given location and the wind climate in both the near and far fields can provide two inspirations. The first inspiration is from wind to waves that the spectral wave climate at any given location can be reconstructed from the corresponding wind projection. Consistent observational directional wave spectra with a record length of more than 30 years are still not available as far as we know, but observational sea surface wind field has a long-term accumulation all over the ocean. Statistical methods can be used to build an empirical parametric model predicting the wave spectra from wind projections over the period when both types of observational data exist. Then the historical directional wave spectra can be estimated from the historical observational wind data. Although such historical spectral wave climates are also available from the hindcast of numerical wave models, some dynamic processes are not fully taken into account or not well represented in numerical models. This type of parametric models can at least supplement the modeled results. Studies have been done using historical SLP data to reconstruct the wave spectra (e.g., *Espejo et al.* 2014; *Rueda et al.* 2017), which should also be able to be conducted using the historical wind information. The second inspiration is from waves to wind that the wind climate can be monitored through the observation of spectral wave climate. Remote sensing and in situ observations of sea surface winds are available globally, but the observation of high-speed wind is still a challenge. The waves generated by high winds will have intermediate or relatively low energy after propagating over large distances which can be observed with better accuracy. Thus, the observational wave spectra can also supplement the observational wind field data, although it might be hard to directly establish a quantitative method to predict the remote wind information for local wave spectra due to the integral effect along wave propagation. Today, more and more data of observational wave spectra with better quality are available from buoys, ground wave radars, and space-borne synthetic aperture radars, which is promising for a better understanding of global wind and wave climates.

**Acknowledgments**

The ERAI data are downloaded by ECMWF Web API. This work is jointly supported by the National Key Research and Development Program of China (Grant No. 2017YFC1404700), the National Natural Science Foundation of China (No. 41806010), and the Discipline Layout Project for Basic Research of Shenzhen Science and Technology Innovation Committee (Grant No. 20170418). HJ and LM are also supported by the Fundamental Research Funds for the Central Universities, China University of Geosciences (Wuhan) (No. CUG170673), and the Guangdong Special Fund Program for Marine Economy Development (Grant No. GDME-2018E001), respectively. The two anonymous reviewers are greatly appreciated for their helpful comments and suggestions.

# Figure Captions

**FIG. 1.** Regions with potential impacts on wave conditions at the selected point (the cross) considering only the along-great-circle propagation and land blocking of waves: (a) distances from the impacting regions to the target point; (b) azimuth from the target point to different locations in the ocean; (c) azimuth from different locations to the target point. Regarding the convention of directions, $0°$ refers to the direction propagating to the north and $90°$ refers the direction propagating to the east. The four wave directions analyzed in Figure 2 are indicated with arrows of corresponding colors in subplot (a).

**FIG. 2.** Time series of monthly averaged wave spectral densities for different frequencies and directions in the selected location over 1981-2000. Each subplot denotes a certain frequency: (a) 0.05 Hz, (b) 0.08 Hz, (c) 0.13 Hz, and (d) 0.21 Hz. The colors denote the directions corresponding to the arrows in Figure 1(a). It is noted that the Y-axes in different subplots are at different scales.

**FIG. 3.** The geographical distributions of (a) maximum cross correlation coefficients between the wave spectrum and wind projections for Point X (the cross), and the distributions of (b) time lag, (c) spectral frequency, and (d) spectral direction corresponding to the maximum correlation. Coefficients not significant at the 99% confidence level are shaded by slashes.

**FIG. 4.** The (a) annual mean wave spectra (with the four wave climate systems marked in red letters), (b) correlations with local winds, (c) spectral standard deviations, and (d) amplitudes of annual cycle for the selected point. A direction of $0°$ corresponds to the wave energy propagating towards the North (for all coordinates of wave spectra in this paper).

**FIG. 5.** The geographical distributions of (a) annual mean wind projections and (b) wind projection STD for Point X (the cross)

**FIG. 6.** Seasonal mean wave spectra for the selected point in (a) DJF, (b) MAM, (c) JJA, and (d) SON. It is noted that the color scales are different for different subplots. The five identified wave systems are marked with red letters in subplot (a) and (c). A direction of 0 ° corresponds to the wave energy propagating towards the North.

**FIG. 7.** The geographical distributions of seasonal mean wind projections for Point X (the cross) in (a) DJF, (b) MAM, (c) JJA, and (d) SON.

**FIG. 8.** Correlation coefficients of the monthly-averaged wave spectra with the (a) SOI, (b) AAOI, and (c) AOI at Point X. Coefficients not significant at the 99% confidence level are shaded by slashes.

**FIG. 9.** Geographical distributions of correlation coefficients of the monthly-averaged wind projections with the (a) SOI, (b) AAOI, and (c) AOI for Point X (the cross). Coefficients not significant at the 99% confidence level are shaded by slashes.

**FIG. 10.** Linear trends for (a) spectral densities and (b) wind projections for Point X over 1992-2017 with the trends not significant at the 95% confidence level shaded by slashes.

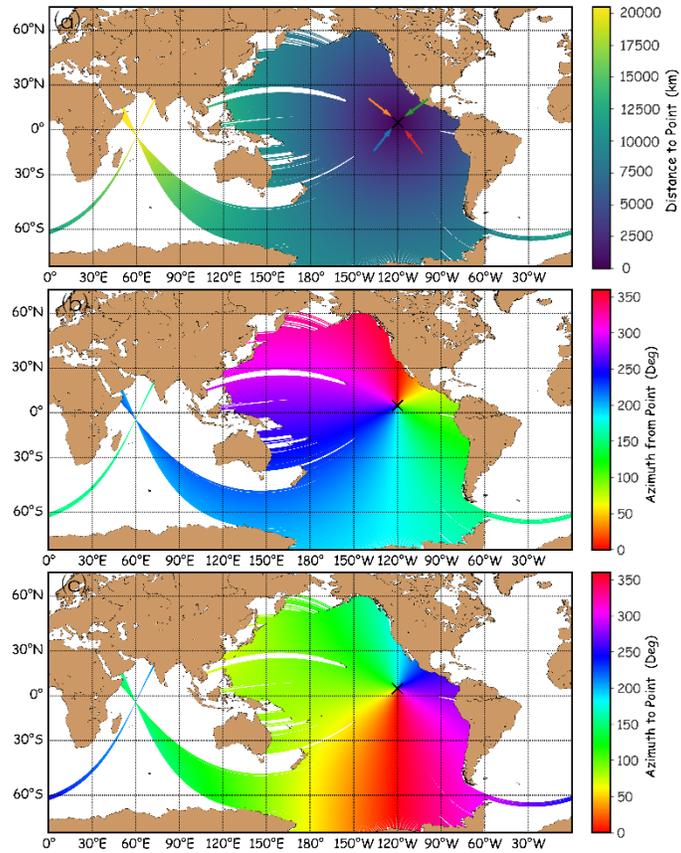

**FIG. 1.** Regions with potential impacts on wave conditions at the selected point (the cross) considering only the along-great-circle propagation and land blocking of waves: (a) distances from the impacting regions to the target point; (b) azimuth from the target point to different locations in the ocean; (c) azimuth from different locations to the target point. Regarding the convention of directions, 0° refers to the direction propagating to the north and 90° refers the direction propagating to the east. The four wave directions analyzed in Figure 2 are indicated with arrows of corresponding colors in subplot (a).

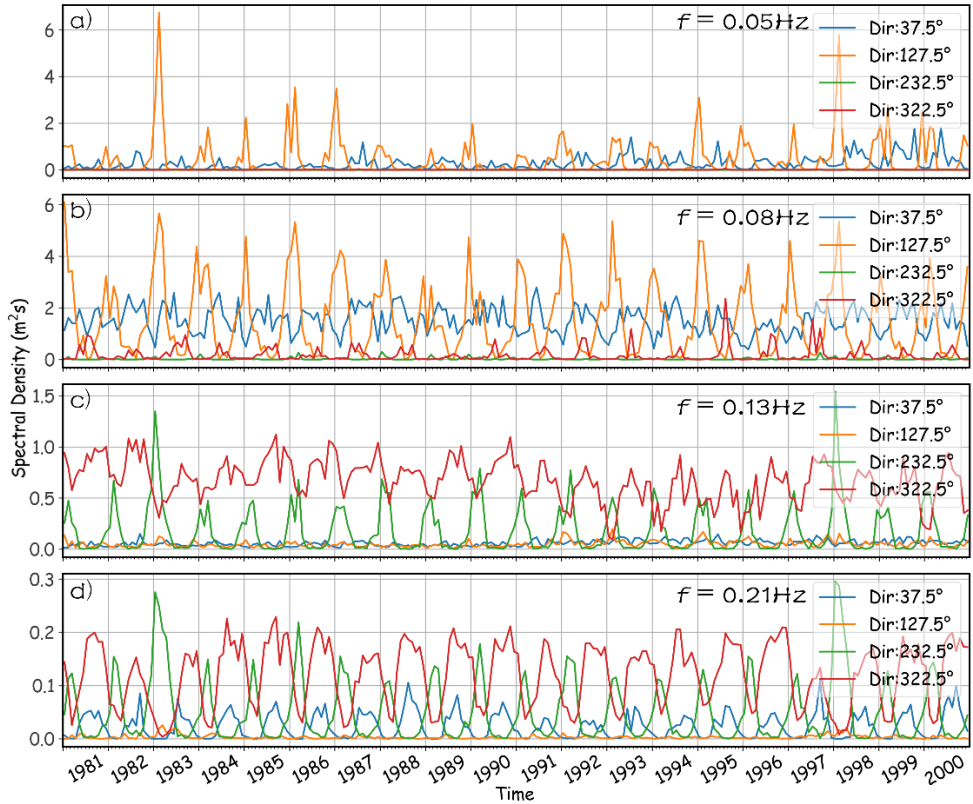

**FIG. 2.** Time series of monthly averaged wave spectral densities for different frequencies and directions in the selected location over 1981-2000. Each subplot denotes a certain frequency: (a) 0.05 Hz, (b) 0.08 Hz, (c) 0.13 Hz, and (d) 0.21 Hz. The colors denote the directions corresponding to the arrows in Figure 1(a). It is noted that the Y-axes in different subplots are at different scales.

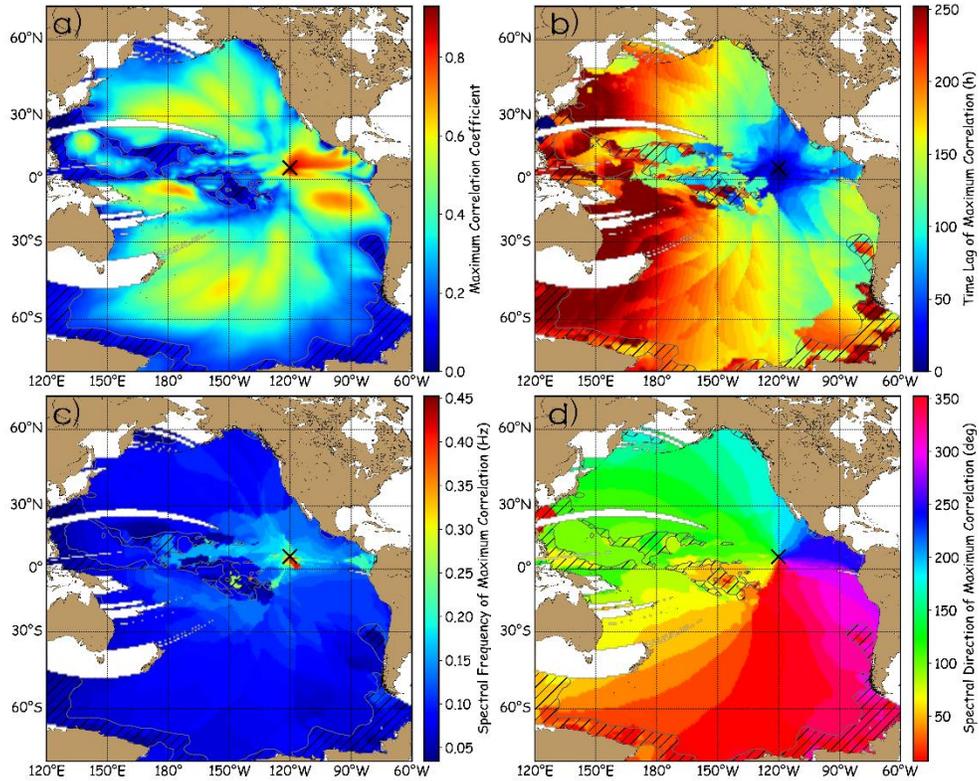

**FIG. 3.** The geographical distributions of (a) maximum cross correlation coefficients between the wave spectrum and wind projections for Point X (the cross), and the distributions of (b) time lag, (c) spectral frequency, and (d) spectral direction corresponding to the maximum correlation. Coefficients not significant at the 99% confidence level are shaded by slashes.

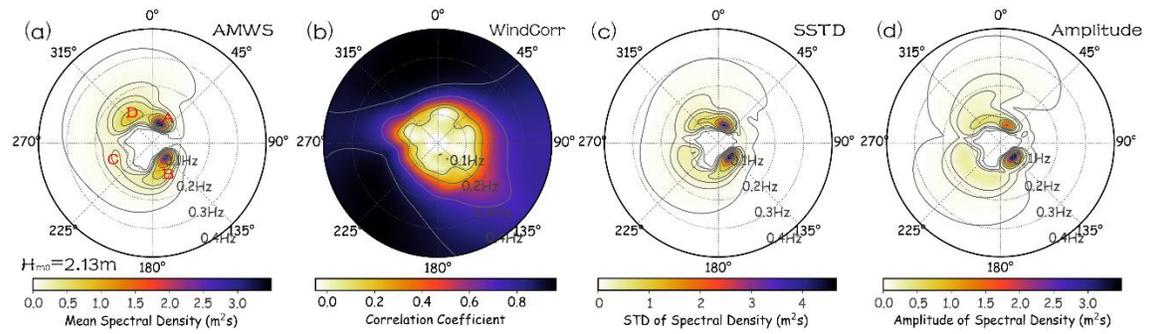

**FIG. 4.** The (a) annual mean wave spectra (with the four wave climate systems marked in red letters), (b) correlations with local winds, (c) spectral standard deviations, and (d) amplitudes of annual cycle for the selected point. A direction of 0 ° corresponds to the wave energy propagating towards the North (for all coordinates of wave spectra in this paper).

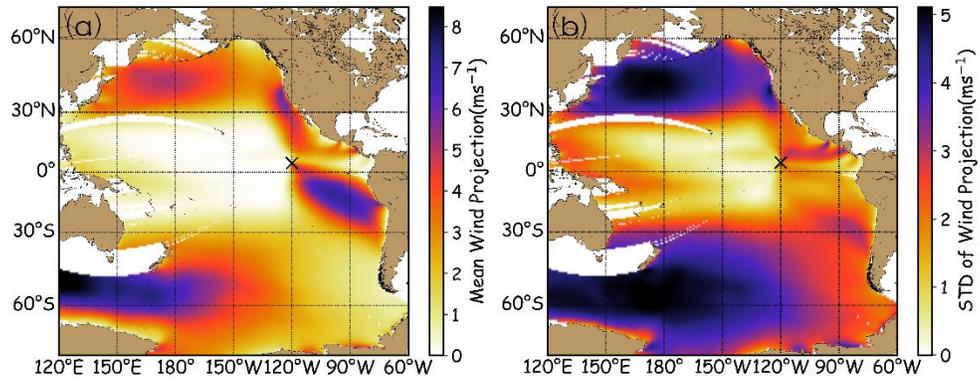

**FIG. 5.** The geographical distributions of (a) annual mean wind projections and (b) wind projection STD for Point X (the cross)

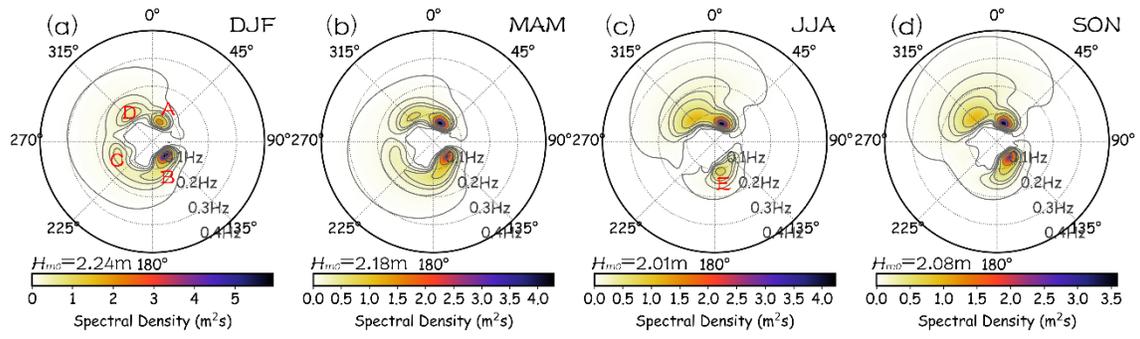

**FIG. 6.** Seasonal mean wave spectra for the selected point in (a) DJF, (b) MAM, (c) JJA, and (d) SON. It is noted that the color scales are different for different subplots. The five identified wave systems are marked with red letters in subplot (a) and (c). A direction of 0 ° corresponds to the wave energy propagating towards the North.

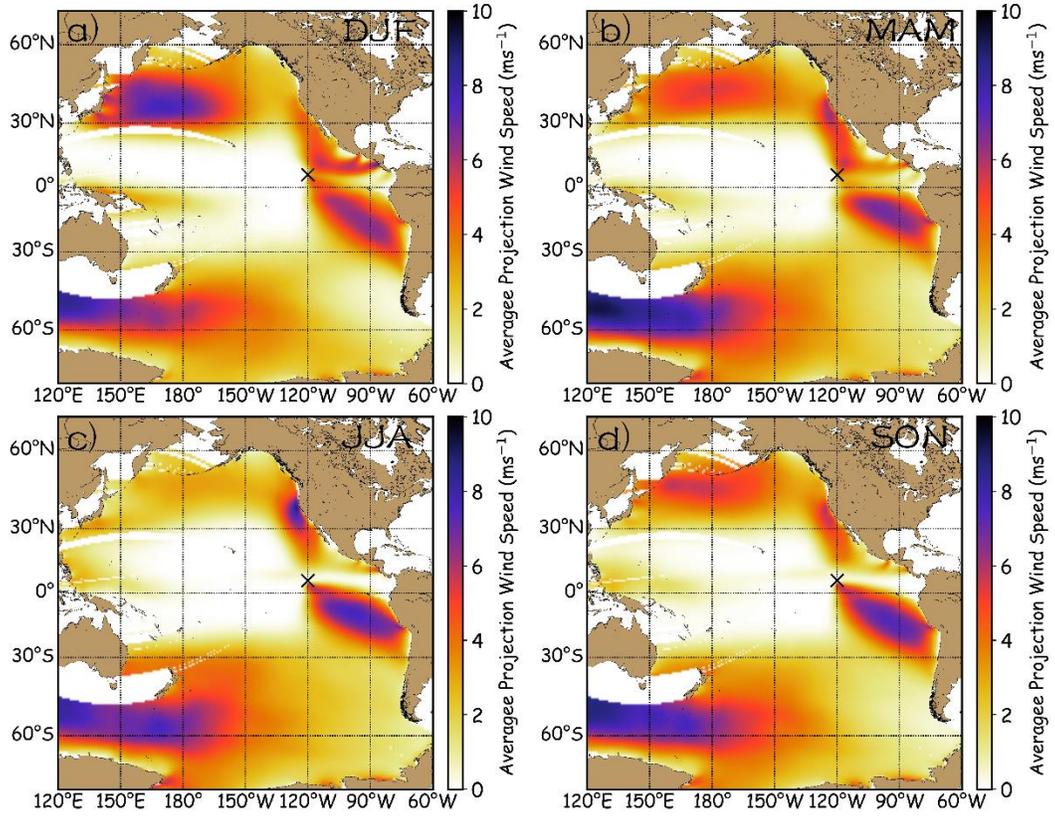

**FIG. 7.** The geographical distributions of seasonal mean wind projections for Point X (the cross) in (a) DJF, (b) MAM, (c) JJA, and (d) SON.

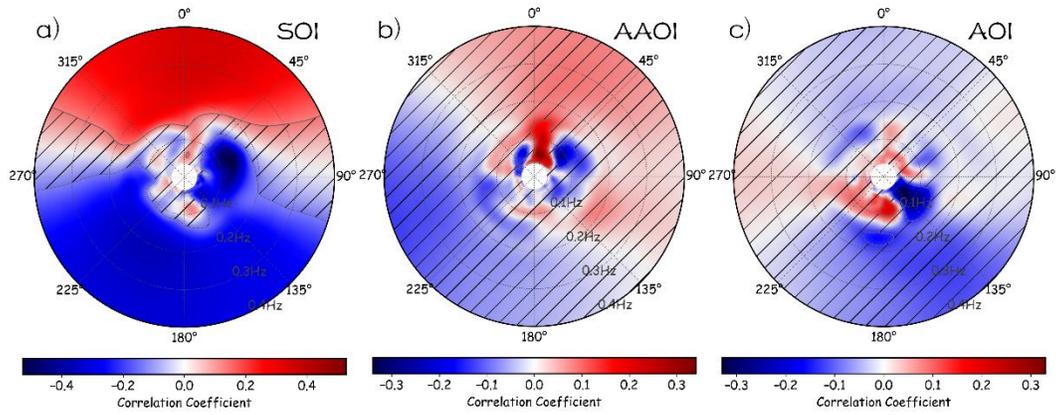

**FIG. 8.** Correlation coefficients of the monthly-averaged wave spectra with the (a) SOI, (b) AAOI, and (c) AOI at Point X. Coefficients not significant at the 99% confidence level are shaded by slashes.

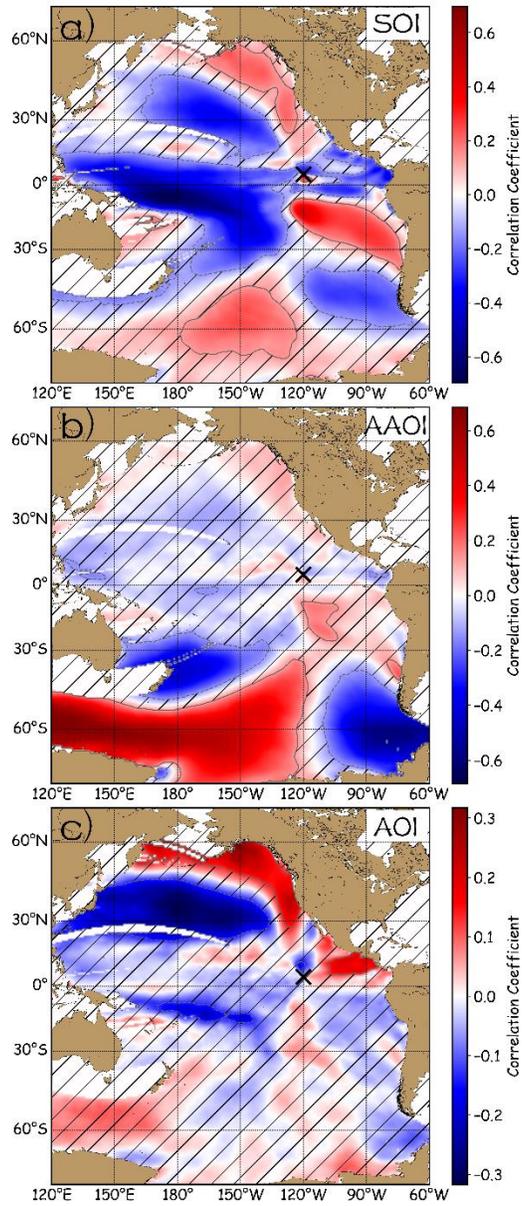

**FIG. 9.** Geographical distributions of correlation coefficients of the monthly-averaged wind projections with the (a) SOI, (b) AAOI, and (c) AOI for Point X (the cross). Coefficients not significant at the 99% confidence level are shaded by slashes.

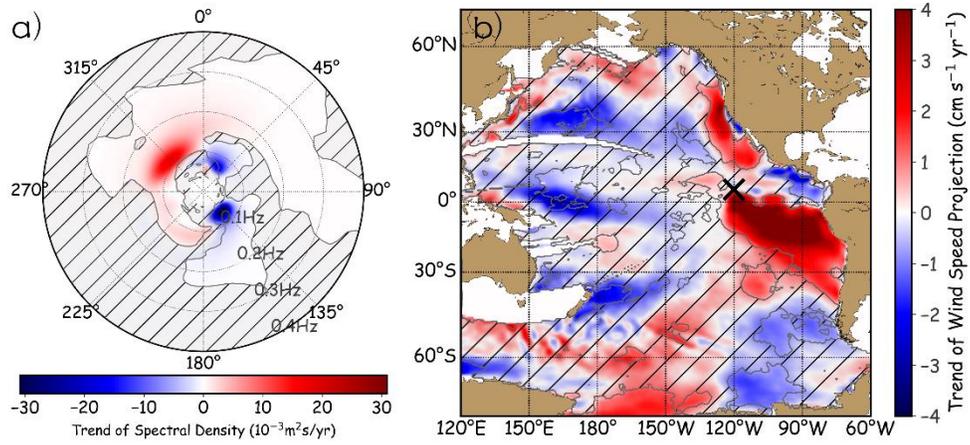

**FIG. 10.** Linear trends for (a) spectral densities and (b) wind projections for Point X over 1992-2017 with the trends not significant at the 95% confidence level shaded by slashes.